\documentclass[conference]{IEEEtran}
\IEEEoverridecommandlockouts
\usepackage{cite}
\usepackage{amsmath,amssymb,amsfonts}
\usepackage{algorithmic}
\usepackage{graphicx}
\usepackage{textcomp}
\usepackage{xcolor}
\usepackage{booktabs}
\usepackage{adjustbox}
\def\BibTeX{{\rm B\kern-.05em{\sc i\kern-.025em b}\kern-.08em
    T\kern-.1667em\lower.7ex\hbox{E}\kern-.125emX}}
\begin{document}

\title{Classification of Carotid Plaque with Jellyfish Sign Through Convolutional and Recurrent Neural Networks Utilizing Plaque Surface Edges}

\author{\IEEEauthorblockN{Takeshi Yoshidomi$^{*}$, Shinji Kume$^{\dagger}$, Hiroaki Aizawa$^{*}$ and Akira Furui$^{*}$}
\IEEEauthorblockA{
Email: \{takeshiyoshidomi, kume, hiroaki-aizawa, akirafurui\}@hiroshima-u.ac.jp\\
$^{*}$Graduate School of Advanced Science and Engineering, Hiroshima University, Higashi-hiroshima, Japan.\\
$^{\dagger}$Department of Clinical Support, Hiroshima University Hospital, Hiroshima, Japan.
}
}

\maketitle

\begin{abstract}
In carotid arteries, plaque can develop as localized elevated lesions.
The Jellyfish sign, marked by fluctuating plaque surfaces with blood flow pulsation, is a dynamic characteristic of these plaques that has recently attracted attention.
Detecting this sign is vital, as it is often associated with cerebral infarction. 
This paper proposes an ultrasound video-based classification method for the Jellyfish sign, using deep neural networks. 
The proposed method first preprocesses carotid ultrasound videos to separate the movement of the vascular wall from  plaque movements. 
These preprocessed videos are then combined with plaque surface information and fed into a deep learning model comprising convolutional and recurrent neural networks, enabling the efficient classification of the Jellyfish sign.
The proposed method was verified using ultrasound video images from 200 patients.
Ablation studies demonstrated the effectiveness of each component of the proposed method.
\end{abstract}


\section{Introduction}
In the carotid arteries which supply blood to the head, localized bulges, known as plaques can form due to factors like hypertension and aging. 
The rupture of these plaques can lead to the spillage of contents into the vascular lumen, potentially causing cerebral infarction by blocking cerebral blood vessels.
Plaques are classified into stable and unstable types, based on characteristics such as the thickness of the fibrous cap and the proportion of lipids they contain, with unstable plaques posing a higher risk of rupture. 
Consequently, early detection of unstable plaques is crucial.

Carotid ultrasound is considered an effective method for examining plaques in the carotid arteries. 
This examination is convenient and minimally burdensome for the patient, enabling real-time video imaging.
It excels in evaluating the dynamic characteristics of unstable plaques. 
Research has been conducted on the quantitative assessment of plaques using ultrasound images and videos~\cite{baroncini2006ultrasonic}.
Additionally, methods based on deep learning are being actively researched.
The effectiveness of these methods has been demonstrated in plaque region segmentation using ultrasound images~\cite{JAIN2021104721} and in the assessment of plaque vulnerability using ultrasound videos~\cite{Guange047528}.

Recently, the \textit{Jellyfish sign}~\cite{kume2010vulnerable}, a dynamic characteristic of unstable plaques, has garnered considerable attention. 
Named for its resemblance to the contraction and expansion of Jellyfish, the Jellyfish sign refers to the phenomenon where the surface of a plaque undulates due to arterial pulsation.
Plaques exhibiting the Jellyfish sign, termed Jellyfish plaques, possess an extremely high risk of rupture, making their early detection crucial.
However, the current assessment of the Jellyfish sign relies on qualitative and subjective visual evaluations.
Therefore, an automated method for classifying the Jellyfish sign from ultrasound videos would be beneficial.

This paper proposes a method for automatically classifying the Jellyfish sign using video analysis and deep learning. 
The Jellyfish sign is characterized by periodic undulations in a specific area of the plaque surface, which necessitates considering this feature for accurate assessment. 
Our proposed method applies preprocessing to the captured ultrasound videos to distinguish the movement of the entire vascular wall from that of the plaque. 
Subsequently, the preprocessed video images are combined with plaque surface information specified by the sonographer. 
This combined data is then input into a deep neural network (DNN) consisting of a convolutional neural network (CNN) and bidirectional long short-term memory (BiLSTM)~\cite{Graves2005}.
This setup enables the classification of the Jellyfish sign, considering the pulsatile changes in the plaque surface.

\section{Jellyfish sign}

The Jellyfish sign is a phenomenon in which the surface of a plaque undulates due to the pulsatile blood flow pressure. 
Reports indicate that 54.8\% of cases exhibiting the Jellyfish sign have developed cerebral infarction~\cite{kume2010vulnerable}, and once a cerebral infarction occurs, there is a likelihood of recurrent strokes within a short period~\cite{kume2010vulnerable}. 
Thus, it is advisable to consider surgery promptly for cases exhibiting the Jellyfish sign, underscoring the critical need for its early detection.

The characteristic feature of the Jellyfish sign is the periodic movement synchronized with the arterial pulsation. 
As the pulsating blood flow exerts pressure on the Jellyfish plaque, the surface of the plaque significantly dips and then rises as the pressure decreases~\cite{Kume2020}. 
The measured ultrasound video captures both the movement of the vascular wall due to pulsation and respiration and the movements of the surface of the Jellyfish plaque. 
Moreover, pulsatile surface undulations are also observed in stable plaques~\cite{Kume2020}, thus focusing on the magnitude of the temporal changes on the plaque surface is necessary to classify the Jellyfish sign.
From these characteristics, it is inferred that correctly evaluating the Jellyfish sign requires consideration of the following:
\begin{itemize}
    \item Separating the movements of the vascular wall due to pulsation and respiration from the movements of the Jellyfish sign.
    \item Focusing on the plaque surface.
    \item Capturing the temporal undulations of the plaque surface.
\end{itemize}
The following section introduces the proposed Jellyfish sign classification method incorporating these key considerations.

\section{Proposed Method} 

\begin{figure}[t] 
  \centering
  \includegraphics[width=0.81\hsize]{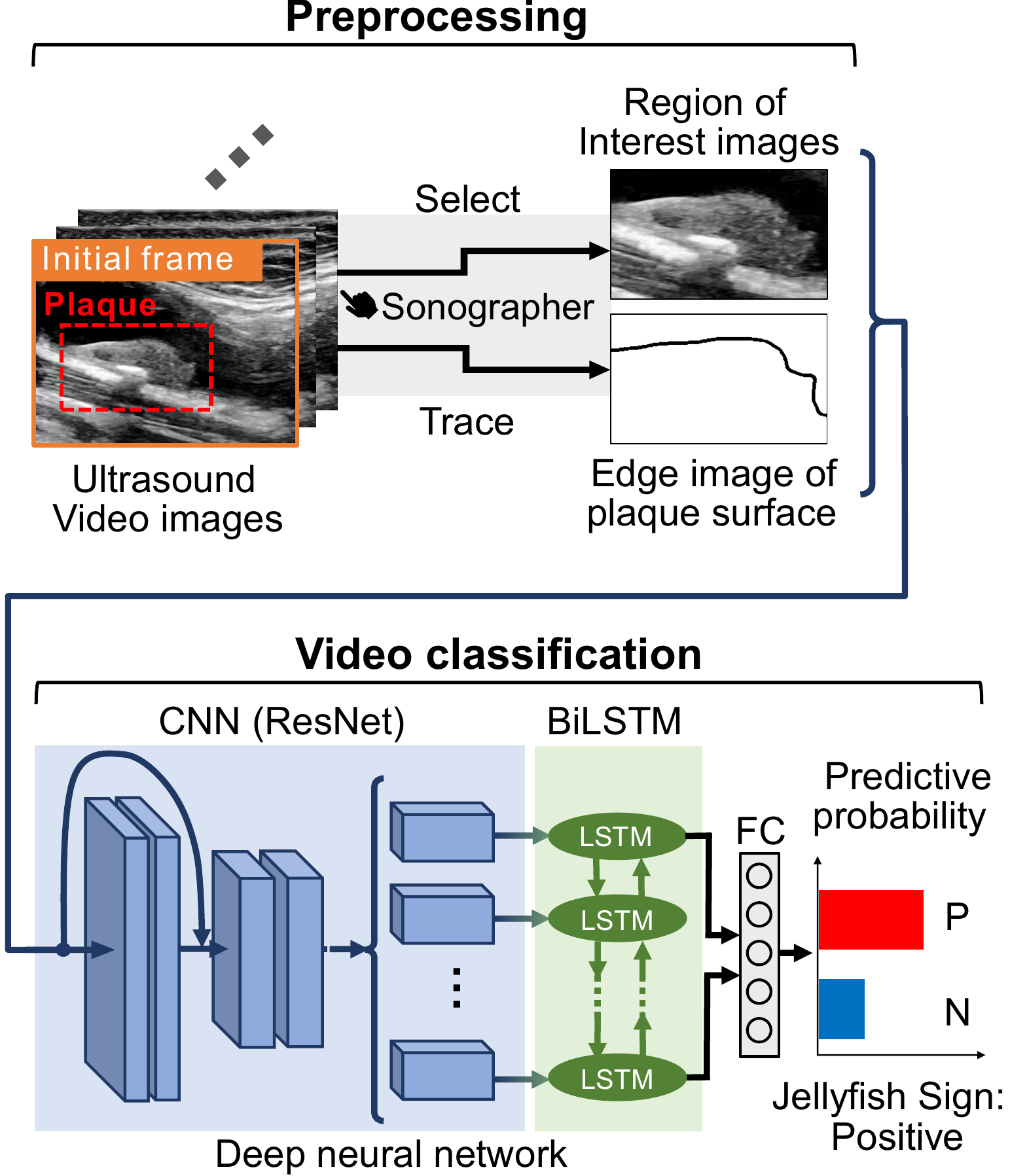}
  \caption{Overview of the proposed method}
  \label{fig:Overview}
\end{figure}

Fig.~\ref{fig:Overview} shows an overview of the proposed method.
The proposed method consists of two main processing stages: preprocessing and video classification. 
In the preprocessing stage, the carotid ultrasound video images undergo preprocessing to extract the plaque movement. 
Subsequently, the preprocessed video images, along with the plaque surface images, are input into the video image classification model to predict the presence or absence of the Jellyfish sign.

\subsection{Preprocessing}\label{AA}

During a carotid artery ultrasound examination, grayscale ultrasound video images (B-mode) are captured. 
The sonographer manually sets the region of interest (ROI) for the initial frame as the area to be examined for the Jellyfish sign. 
The sonographer can arbitrarily set the size of the ROI, but its aspect ratio is fixed at 5:3. 
Furthermore, by tracing the surface of the plaque within the ROI at the initial frame, an edge image, referred to as the plaque surface image, is created to emphasize the boundary of the plaque surface.
The plaque surface image is binarized, making the edges black and the remaining area white.

In the preprocessing stage, template matching is applied to suppress the movements of the vascular wall caused by pulsation and respiration in the captured video images. 
This aims to extract only the movement of the plaque surface. 
The ROI image is set as the template $\mathcal{T}$.
For each frame within the captured video, the area where the normalized cross-correlation function 
\begin{equation}
R(i,j) = \frac{\sum_{a,b}\mathcal{T}(a,b)\mathcal{I}_t(i+a, j+b) }{\sqrt{\sum_{a,b}\mathcal{T}(a,b)^2}\sqrt{\sum_{a,b}\mathcal{I}_t(i+a,j+b)^2}}
\end{equation}
is maximized is extracted.
Here, $\mathcal{I}_t(i, j)$ represents the pixel value of the captured video image at frame $t$ and coordinates $(i, j)$. 
This process is applied to all frames to construct the plaque video image. 
Finally, as the size of the ROI varies in the measured video, both the plaque video image and plaque surface image are resized to $W \times H$ (pixels).

\begin{figure}[t] 
  \centering
  \includegraphics[width=0.575\hsize]{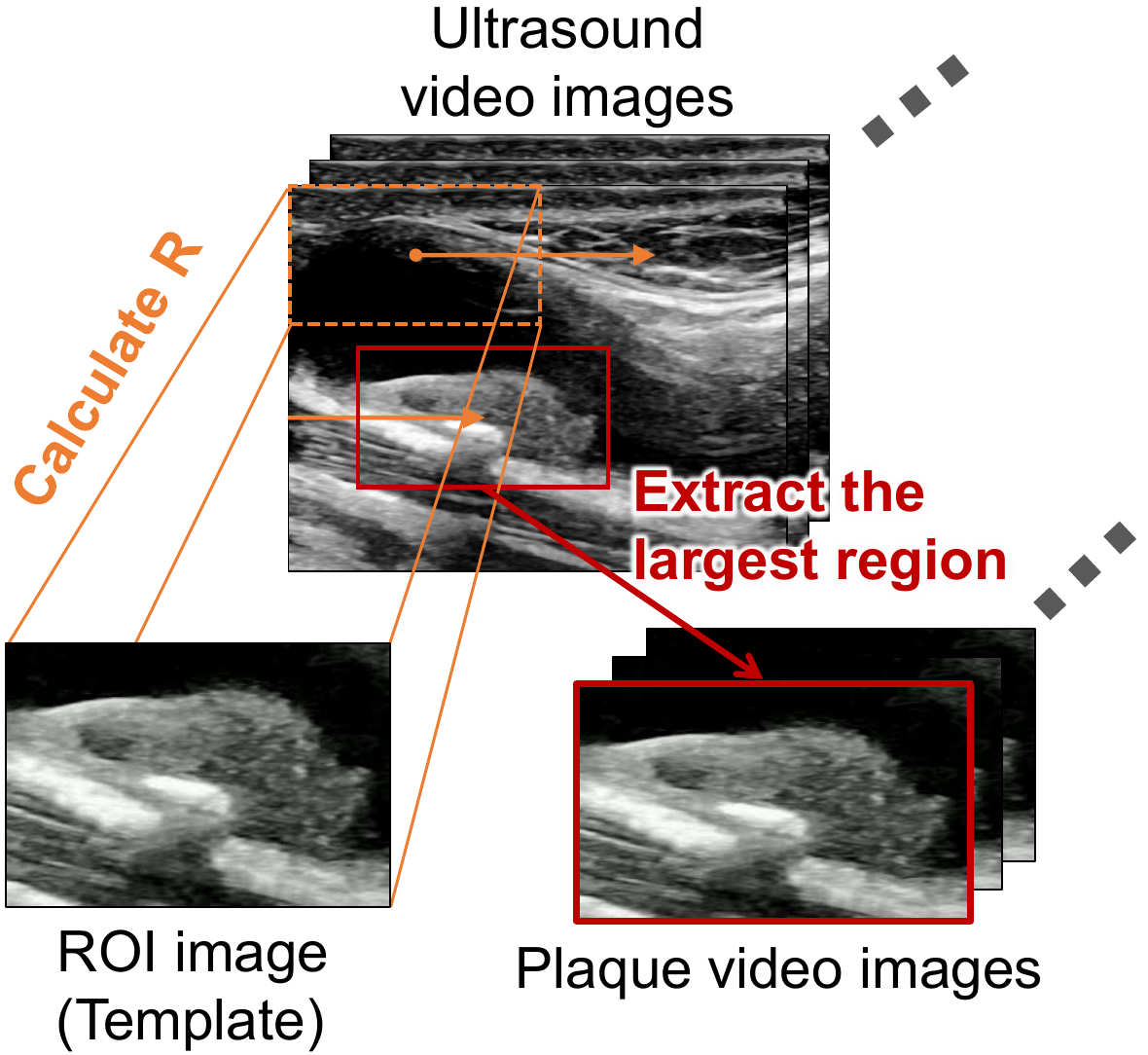}
  \caption{Schematic diagram of the template matching process}
  \label{fig:template}
\end{figure}

\subsection{Plaque Video Classification with CNN-BiLSTM}

\subsubsection{Input structure}
In this paper, to facilitate classification focused on the plaque surface movement, the plaque surface image in the initial frame is utilized. 
Since only a single plaque surface image is obtained from the initial frame, this image is duplicated and concatenated with each frame of the corresponding plaque video sequence.
Thus, the surface image does not exactly match the plaque surface's position throughout the video but provides a general representation of the surface area for analysis.
Specifically, a two-channel video image $\mathbf{x}_t \in \mathbb{R}^{W \times H \times 2}$ ($t = 1, \ldots, T$, where $T$ is the length of the video) is constructed by concatenating the plaque video image and the plaque surface image along the channel dimension. 
These concatenated video images are used as inputs in the classification model.

\subsubsection{Model architecture}
For the classification model, a DNN composed of a CNN and BiLSTM is used. 
The CNN is a feedforward network with two characteristic layers: convolutional and pooling layers, which extract spatial features from the input video images. 
In the proposed method, the CNN architecture is a residual network (ResNet)~\cite{he2016deep}, equipped with residual blocks based on skip connections, enabling stable learning even with deeper layers. 
Specifically, an 18-layer ResNet is used in this paper.

The BiLSTM, a type of recurrent neural network, connects to the latter part of the CNN to perform feature extraction, considering temporal dependencies from the spatial features extracted by CNN. 
It combines LSTM~\cite{hochreiter1997long}, capable of processing long-term dependencies, in both directions, allowing for learning that considers all preceding and following information in the data. 
The output of the BiLSTM is passed through a fully connected (FC) layer with a ReLU function, followed by another FC layer with a sigmoid function. 
Finally, the model predicts the probability of a positive Jellyfish sign at the final time step $T$.

\subsubsection{Learning algorithm}
Given $\{(\mathbf{x}^{(n)}_t, y^{(n)})\}_{n=1}^N$ as the $N$ labeled training dataset, consider the network training. 
Here, $y^{(n)} \in \{0, 1\}$ is a binary label indicating the presence or absence of the Jellyfish sign.
For the loss function, we use the following cross-entropy loss:
\begin{equation}
\mathcal{L} = -\frac{1}{N}\sum^N_{n=1}\{y^{(n)}\ln \hat{y}^{(n)}+(1-y^{(n)})\ln (1-\hat{y}^{(n)})\},
\end{equation}
where $\hat{y}^{(n)}=p(y^{(n)}|\mathbf{x}_T^{(n)})$ represents the predicted probability output by the model.
To minimize this loss function, the weight parameters of the CNN and BiLSTM are updated collectively based on the backpropagation through time.

\section{Experiments}

To validate the effectiveness of the proposed method, we conducted a Jellyfish sign classification experiment. 
In this experiment, we utilized a dataset of ultrasound video images (B-mode) obtained from carotid ultrasound examinations conducted at Hiroshima University Hospital. 
All measurements were conducted in accordance with the Helsinki Declaration, with the approval of the Ethics Committee of Hiroshima University (Approval Numbers: E-585 and E-721), and with informed consent obtained from the patients.

\subsection{Experimental Setup}

\subsubsection{Dataset}
The dataset comprised 200 cases, including 100 positive and 100 negative cases for the Jellyfish sign.
The video frame rate was 30 fps, with a length of 84 $\pm$ 18 frames. 
The size of the ROI was set to a width of 238 $\pm$ 68 pixels and a height of 142 $\pm$ 41 pixels. 
Each ultrasound video in the dataset was pre-extracted by a sonographer to include data for two heartbeats. 
The Jellyfish sign labels were assigned to each video based on annotations by the sonographer. 
Although annotations of the Jellyfish sign lesion areas were also made by the sonographer, this information was not used in the analysis presented in this paper.
To evaluate the classification performance, a 5-fold cross-validation was applied to the dataset, dividing it into training and testing data. 
Additionally, a random 10\% of the training data was set aside as validation data.

\subsubsection{Implementation details}
In the preprocessing step, the resize settings were $W = 224$ pixels and $H = 134$ pixels. 
For training the proposed method, the ResNet, placed at the beginning of the model, was first pre-trained using the ImageNet dataset~\cite{deng2009imagenet}, and then the entire network was retrained with our Jellyfish sign dataset. 
The network training employed stochastic gradient descent with a batch size of 16 and a learning rate of 0.01. 
To prevent overfitting, spatial data augmentation was applied to the training data by randomly flipping images horizontally or vertically with a 50\% probability.
Additionally, temporal data augmentation was performed through temporal random cropping using a window of 45 frames (i.e., $T = 45$). 
The maximum number of epochs was set to 150, and the model instance with the minimum validation loss at that epoch was used for the final evaluation of the test data.

\subsubsection{Evaluation metrics}
To evaluate the classification performance on the test data, accuracy, precision, and recall metrics were used. 
Each metric was calculated three times with different random seeds for initialization, and the averages of these metrics were reported.

\subsubsection{Ablation study}
To analyze how each element of the proposed method impacts its overall effectiveness, a two-perspective ablation study was carried out.
First, to confirm the usefulness of combining plaque surface images with plaque video images as input, the performance of the method using only plaque video images as input to our video classification model was evaluated. 
Second, to investigate the impact of different approaches to handling temporal information, the performance was evaluated when our video classification model (CNN-BiLSTM) was replaced with the following two configurations:
\begin{itemize}
    \item \textbf{CNN-LSTM:} A model using LSTM rather than BiLSTM, which can only consider forward temporal sequences.
    \item \textbf{CNN-FC:} A model where FC layers are directly connected after the CNN, incapable of considering temporal dynamics.
\end{itemize}

\subsection{Results and Discussion}\label{RD}

\begin{figure*}[t] 
  \centering
  \includegraphics[width=0.87\hsize]{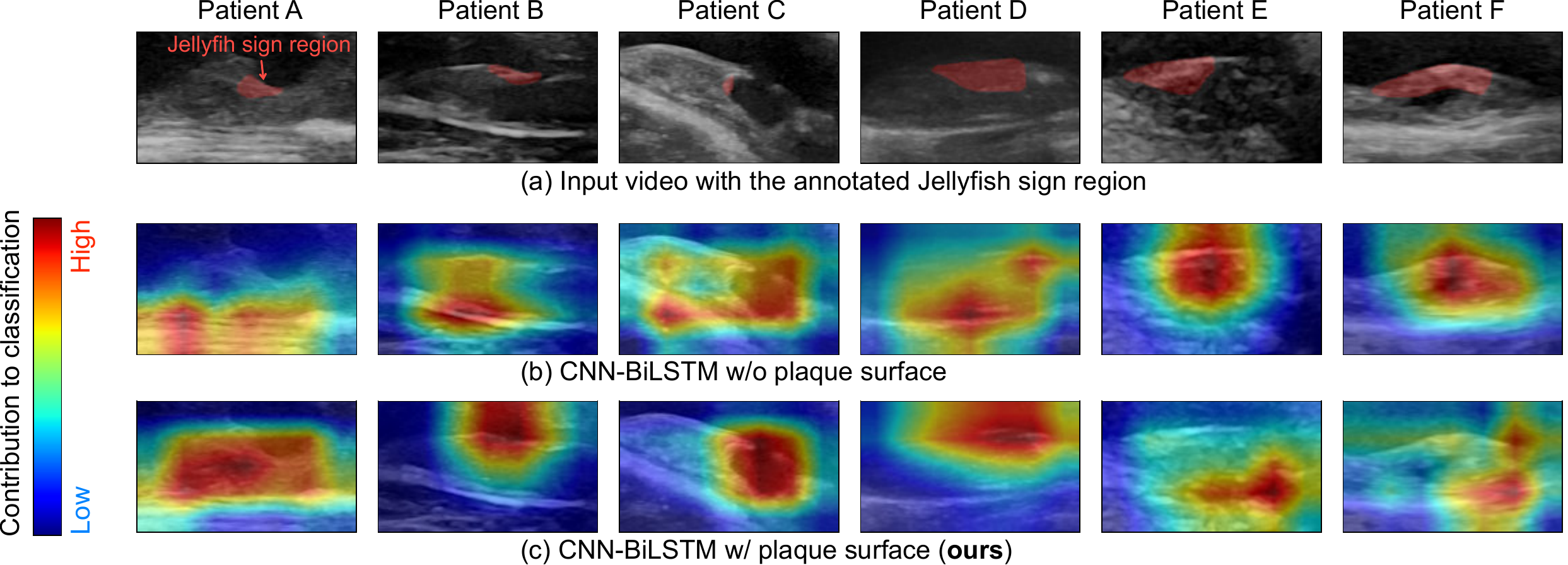}
  \caption{Visualization of activation maps using Grad-CAM++. The top row displays the initial frame of the ultrasound video images used as input. The semi-transparent red area in the image marks the region annotated by sonographer as the lesion area of the Jellyfish sign. In the activation maps, areas with more intense red indicate a more considerable contribution to the classification.}
  \label{fig:GradCam}
\end{figure*}

\subsubsection{Effectiveness of plaque surface images}
Table~\ref{table:edge} compares the performance of the proposed method, which uses a combination of plaque surface images and plaque video images as input, with a method that only uses plaque video images.
The proposed method outperformed the latter across all metrics, indicating that explicitly including plaque surface images in the input more effectively extracts the characteristics of the Jellyfish sign.

Fig.~\ref{fig:GradCam} shows typical examples of class activation maps for Jellyfish sign-positive cases using Grad-CAM++~\cite{chattopadhay2018grad}, applied to the last convolutional layer of ResNet. 
This visualization technique highlights areas contributing to positive classification.
In the proposed method with plaque surface images, areas close to the plaque surface, particularly around the actual Jellyfish sign, are more frequently and strongly activated (Patients A--D).
This implies that adding plaque surface images to input data enhances feature extraction focused on the plaque surface. 
However, there were instances where strong activation occurred in areas other than the plaque surface (Patients E and F), even with plaque surface images included. 
In these cases, not only the plaque surface but also the internal moving parts of its structure tended to become active.
This suggests that important motion information for classifying the Jellyfish sign may exist beyond the plaque surface, but further analysis is required to elucidate this.

\subsubsection{Effectiveness of temporal information}
Table \ref{table:temporal} presents the results of the ablation study on the classification model. 
When temporal information was not used (CNN-FC), the classification performance was the lowest, indicating the importance of learning temporal information in Jellyfish sign classification. 
Furthermore, the proposed method with BiLSTM outperformed the CNN-LSTM model, likely because the bidirectional structure allows for more effective capture of the periodic movement of the Jellyfish sign by considering the temporal sequence of data from both forward and backward directions.
From these results, the DNN with recurrent structure can realize the classification of the Jellyfish sign, which is a dynamic characteristic of plaques that have been conventionally identified through visual examination.

\begin{table}[t]
\centering
  \caption{Results on the effect of plaque surface images}
  \label{table:edge}
  \scalebox{0.93}{
  \begin{tabular}{@{}l|ccc@{}}
  \toprule
    Input structure &  Accuracy & Precision & Recall \\
    \midrule
    w/o plaque surface & $0.778 \pm 0.006$ & $0.785\pm 0.004$ & $0.773\pm 0.006$   \\
    w/ plaque surface (\textbf{ours}) & \scalebox{0.94}[1.0]{$\mathbf{0.805\pm 0.011}$} & \scalebox{0.94}[1.0]{$\mathbf{0.810\pm 0.008}$} & \scalebox{0.94}[1.0]{$\mathbf{0.802\pm 0.011}$} \\
    \bottomrule
  \end{tabular}
  }
\end{table}

\begin{table}[t]
  \caption{Results on the effect of temporal information}
  \label{table:temporal}
  \centering
  \scalebox{0.96}{
  \begin{tabular}{@{}l|ccc@{}}
  \toprule
     Classification model &  Accuracy & Precision & Recall 
     \\
  \midrule
    CNN-FC & $0.775 \pm 0.011$ & $0.780 \pm 0.008$ & $0.771 \pm 0.012$   \\
    CNN-LSTM & $0.787 \pm 0.030$ & $0.800 \pm 0.020$ & $0.785 \pm 0.035$  \\
    CNN-BiLSTM (\textbf{ours}) & \scalebox{0.94}[1.0]{$\mathbf{0.805\pm 0.011}$} & \scalebox{0.94}[1.0]{$\mathbf{0.810\pm 0.008}$} & \scalebox{0.94}[1.0]{$\mathbf{0.802\pm 0.011}$}   \\
  \bottomrule
  \end{tabular}
  }
\end{table}

\section{Conclusion}
This paper proposed a method for automatically classifying the Jellyfish sign in carotid artery plaques. The method combines ultrasound video images with plaque surface images and inputs them into a DNN, enabling the classification of the Jellyfish sign while considering its features.

An experiment with a dataset of 200 cases shows that the proposed method achieved a classification accuracy of 80.5\% in classifying the Jellyfish sign.
We discovered that key contributions to this performance include incorporating plaque surface images and using a bidirectional recurrent structure. 
Future work will focus on refining the pre-training dataset and enhancing feature extraction through attention mechanisms and the integration of electrocardiogram data.

\bibliographystyle{IEEEtran}
\bibliography{reference} 

\end{document}